\documentstyle[12pt]{article}
\begin{document}
\renewcommand{\baselinestretch}{2}
\newcommand{\noi}{\noindent}
\newcommand{\be}{\begin{equation}}
\newcommand{\ee}{\end{equation}}
\newcommand{\bea}{\begin{eqnarray}}
\newcommand{\eea}{\end{eqnarray}}
\newcommand{\ba}{\begin{array}}
\newcommand{\ea}{\end{array}}
\newcommand{\half}{\frac{1}{2}}
\newcommand{\fourth}{\frac{1}{4}}
\newcommand{\kk}{k_{2}^{2}+k_{3}^{2}}
\newcommand{\km}{k_{1}^{2}+m^{2}}
\newcommand{\pic}{\scriptscriptstyle}
\newcommand{\esp}{\vspace{0.5cm}}

\newcommand{\nS}{\mbox{$n_{S}$}}
\newcommand{\nT}{\mbox{$n_{T}$}}
\newcommand{\xT}{\mbox{$x_{T}$}}
\newcommand{\CT}{\mbox{$C^{T}_{2}$}}
\newcommand{\CS}{\mbox{$C^{S}_{2}$}}
\newcommand{\Hpres}{\mbox{$100h\,{\rm km\,sec^{-1}\,Mpc^{-1}}$}}

\begin{center}
{\Large \bf EFFECTS OF BULK VISCOSITY ON COSMOLOGICAL EVOLUTION}
\end{center}

\vspace{.2in}
\begin{center}
Luis O. Pimentel\footnote{email: lopr@xanum.uam.mx} and Luz M. Diaz-Rivera \\
{\it  Departamento de F\'{\i}sica Universidad Aut\'onoma
Metropolitana-Iztapalapa, Apartado Postal 55-534, M\'exico D. F., MEXICO.}

\vspace{.2in}
\end{center}

\begin{quote}
{\bf Abstract:}

The effect of bulk viscisity on the evolution of the
homogeneous and isotropic cosmological models is considered.
Solutions are found, with a barotropic equation of state,
and a viscosity coefficient that is proportional to a power
of the energy density of the universe. For flat space, power law
expansions, related to extended inflation are found as well as
exponential solutions, related to old inflation; also a solution
with expansion that is an exponential of an exponential of the
time is found.

\end{quote}

\begin{center}
\vspace{.2in}
PACS numbers~~~98.80.Cq, 04.30.-w, 04.80.Nn\\
\end{center}

\vspace{.3in}


\def\square{\mathchoice\sqr54\sqr54\sqr{6.1}3\sqr{1.5}6\,\,}
\def\qa {^{\prime}}
\def\qq {^{\prime\prime}}
\def\qb {^{\prime 2}}
\def\noi {\noindent}
\baselineskip=20pt

\section {Introduction}
The effect of negative pressure and bulk viscosity on the
evolution of homogeneous isotropic cosmological models have been
considered recently by Wolf \cite{Wolf}. He considered a viscosity
coefficient that is proportional to the energy density of the matter
in the universe and flat topology of the three space. In this paper we
extend his consideration to the case
when the viscosity coefficient is proportional to a power of the
energy density and arbitrary curvature of the three space.

We want to consider Eintein's field equations for
the isotropic and homogeneous line
element,
\be
ds^2=dt^2-a^2(t)\left [ {dr^2\over 1-k r^2}+
r^2(d\theta^2+\sin^2\theta d\phi^2) \right ],
\ee

\noi and with the material substratum corresponding to a fluid with
bulk viscosity, whose energy momentum tensor is given by

\be
T_{\mu \nu}= (\rho + {p-\zeta \theta})u_\mu u_\nu+{p-\zeta
\theta}g_{\mu \nu},\quad \theta= u^{\sigma}_{;\sigma}.
\ee

\noi They are

\be
 ({\dot a \over a})^2 + {k \over a^2}  = {8\pi
\rho G\over 3  } ,
\label{eq:rho}
\ee

\be
\frac{2 \ddot a}{a} + ({\frac{\dot a}{ a}})^2 +{k\over a^2}  = -8\pi G
(p-3 \zeta \frac{\dot a }{a}),
\ee

\noi eliminating the energy density from the above equations we obtain,

\be
\frac{2 \ddot a}{a} + ({\frac{\dot a}{ a}})^2 +
{k\over a^2}+({\dot a \over a})^2  = -8\pi G
(p-3 \zeta \frac{\dot a }{a}).
\label{eq:una}
\ee

\noi In order to continue we need some equation of state in the form
$p=p(\rho)$  and $\zeta = \zeta (\rho)$ that we assume of the
following form,

\be
p= \epsilon \rho,\;\;\zeta= \zeta_0 \rho^n.
\ee

The last equation of state has been used before. If n=1 we have a
relativistic fluid, n=3/2 corresponds to a string dominated
universe in the sense of Turok \cite{turok}. Some other values of n arise when
quantum effects are considered. With the above equations of state
Eq.(\ref{eq:una}) reduces to,

\be
2a\ddot a +A({\dot a}^2 +k ) + B \frac{\dot a a }{a^{2n}} ({\dot a}^2 +k
)^{n}=0,
\label{eq:diff}
\ee

\noi where

\be
A=3\epsilon +1 ,\; B= -(8\pi )^{(1-n)} 3 ^{(1+n)} \zeta_0.
\ee

After solving the above equation we use Eq.(\ref{eq:rho}) to calculate the
corresponding energy density of the models.

To solve the differential equation it is useful to
define the following dependent variable,

\be
y=y(a)= {\dot a}^2 +k
\ee

\noi and now our differential equation is

\be
a y'+ A y + B a^{(1-2n)} (y-k)^{1/2}y^n=0.
\label{eq:dife}
\ee

\noi In the next sections we solve exactly this equation for some values of
A, B and n.

\section{Flat Space Solutions}

For the case of flat space (k=0), Eq.(\ref{eq:dife}) reduces to

\be
2a\ddot a +A{\dot a}^2  + B  a^{1-2n}  {\dot a}^{1+2n}=0,
\ee

\be
a y'+ A y + B a^{(1-2n)} y^{1/2+n}=0.
\ee

\subsection{n=1/2}
For n = 1/2 the above equation is linear,

\be
a y'+ C y  = 0,\quad C=A+B=3\epsilon +1  -(8\pi )^{(1/2)} 3 ^{(3/2)}
\zeta_0,
\ee

\noi and we can find a(t) explicitely. For $C=-2$ we have,

\be
a= a_0 e^{H t}, \;{\rm for}\quad  C=-2,(8\pi )^{(1/2)} 3 ^{(3/2)}
\zeta_0=3(\epsilon
+1),
\ee

\noi and the corresponding energy density is

\be
\rho =\rho_c=\frac{3 H^2}{8 \pi G}
\ee

\noi and for $C\ne -2$ the solution is

\be
a= [k_1+k_2 t]^{\frac{2}{C+2}}, \; {\rm for}\;\; C\ne -2
\ee

\noi with the density

\be
\rho =\frac{3 }{8 \pi G} {(\frac{2 k_2}{C+2})}^2{(\frac{1}{k_1 +k_2
t})}^2
\ee

\noi where $k_i$, $a_0$, and H are constants of integrations. The first of
the above solutions is an old inflationary solution, the second
one for any $\zeta_0\ne 0$ will correspond to
the new infationary solutions.

\subsection{$n\ne 1/2,A\ne -2,(\epsilon\ne -1$) }
For $n\ne 1/2$ Eq.(\ref{eq:dife}) is of the Bernoulli type and the solution
for $A\ne -2,(\epsilon\ne -1$), is

\be
y= [y_1 a^{(2n-1)A/2}-\frac{B}{A+2}a^{1-2n}]^{\frac{2}{1-
2n}}, \; A\ne -2\; (\epsilon\ne -1).
\ee

For n=1 it is possible to do the second integration
for arbitrary value of the integration constant $y_1$, the result
is,

\be
2y_1 a^{\frac{3(\epsilon+1)}{2}}+ 9\xi_0\ln (a) =\pm 3(\epsilon +1)( t +t_0)
\ee

\noi with the density given by

\be
\rho=\frac{3}{8\pi G}[y_1 a^{\frac{3(\epsilon +1)}{2}} +\frac{9
\eta_0}{3(\epsilon +1)}]^{-2}.
\ee

This solution has the exponential behaviour for earlier times,
with Hubble parameter $H=( \epsilon +1)/3\xi_0$, and a power law
expansion at late times with exponent equal to $2/3(\epsilon +1)$.
This late behaviour corresponds to the standard cosmology without
viscosity. Therefore this family of solutions provides a natural
end to the inflationary era with exponential expansion
with a transition to a standard cosmological era if $\epsilon \ge -1/3$
or to an extended inflatinary period if $\epsilon < -1/3$.

To proceed for $n \ne 1$ we have to make the choice $y_1 =0$, and we have

\be
y={\dot a}^2= [-\frac{B}{A+2}a^{1-2n}]^{\frac{2}{1-
2n}}, \; A\ne -2\; (\epsilon\ne -1)
\ee

from where the expansion factor can be obtained explicitely,

\be
a(t)=a_0 e^{Ht},\quad H=[\frac{B}{2+A}]^{\frac{1}{1-2n}}
\ee
with the corresponding energy density

\be
\rho =\frac{3 H^2}{8 \pi G}.
\ee

\bigskip

\subsection{$n\ne 1/2,A= -2,(\epsilon= -1$) }

In case our fluid is of the vacuum type, Eq.(\ref{eq:dife}) can
be integrated,

\be
y={\dot a}^2= a^2[y_1 +\frac{(2n-1)B}{2}\ln a]^{\frac{2}{1-
2n}},\quad  A=-2 \;(\epsilon=-1).
\ee

In this case we can do the second integration without any
assumption about the integration constant $y_1$

For $n\ne 0$ the solution is

\be
a(t)=a_0\exp[\frac{2(a_1+nB t)^{\frac{2n-1}{2n}}}{(2n-1)B}],
\ee

\be
\rho = \frac{3 H^2}{2 \pi G} (a_1+n B t)^{\frac{2n-2}{n}}.
\ee

For n=0 the solution is

\be
a(t)=a_0\exp[a_1 \exp[-B t/2]],
\ee

\be
\rho =\frac{3 H^2}{8 \pi G} (\frac{-a_1 B}{2}) \exp[-Bt/2]
\ee

\noindent here $a_i$ are the redefined constants of integration. From
the above solutions in the case of flat space we see that for any
equations of state ( arbitrary $\epsilon$ and n ) we have
inflation, i.e., accelerated expansion. In particular for
$\epsilon =-1$, that gives the old inflation in the usual case
without viscosity, we see that viscosity gives a "superinflation",
 as long as $a_1>0$.

\section{Non-flat Solutions}

In the case of non-flat space Eq.(\ref{eq:dife}) can be solved for
n=0, and A=0 ($\epsilon=-1/3$) the solution is

\be
a(t)= a_0+a_1 e^{-B t/2},
\ee

\be
\rho =\frac{3}{8\pi G}[\frac{k+(\frac{a_1B}{2})^2 e^{-Bt}}{(a_0+a_1 e^{-B
t/2})^2}].
\ee

We notice that the expansion factor does not depend on the value of the
curvature of the three dimensional space. the reason for that can
be seen from equation (7) since for n=A=0 the dependence on the
curvature disappears. Nevertheless, the curvature is present in the
energy density.
We recall here that the equation of state corresponds to a gas of
textures and also to some cases of cosmic strings.
\section {ACKNOWLEDGMENTS}

This work was partially supported by the CONACYT GRANT 1861-E9212.

\end{document}